\documentclass[aps,preprintnumbers,floatfix,article,amsmath,amssymb,floatfix,10pt,prd,superscriptaddress,nofootinbib,twocolumn]{revtex4}
\usepackage{bm}
\usepackage{amsfonts}
\usepackage{latexsym}
\usepackage{graphicx}
\usepackage{amsmath}
\usepackage{palatino}
\usepackage{mathpazo}
\usepackage{textcomp}
\usepackage{float}
\usepackage{booktabs}
\usepackage{dcolumn}
\usepackage{ragged2e}
\usepackage{hyperref}
\hypersetup{colorlinks,citecolor=magenta}
\hypersetup{colorlinks=true,linkcolor=red,filecolor=magenta,urlcolor=blue}
\usepackage{amsmath}
\usepackage{xcolor}
\usepackage{orcidlink}
\usepackage{epsfig}
\usepackage{caption}
\usepackage[caption=false]{subfig}
\def\jnl@style{\it}
\def\aaref@jnl#1{{\jnl@style#1}}

\def\aaref@jnl#1{{\jnl@style#1}}

\def\aj{\aaref@jnl{AJ}}                   
\def\apj{\aaref@jnl{ApJ}}                 
\def\apjl{\aaref@jnl{ApJ}}                
\def\apjs{\aaref@jnl{ApJS}}               
\def\apss{\aaref@jnl{Ap\&SS}}             
\def\aap{\aaref@jnl{A\&A}}                
\def\aapr{\aaref@jnl{A\&A~Rev.}}          
\def\aaps{\aaref@jnl{A\&AS}}              
\def\mnras{\aaref@jnl{Mon.~Not.~Roy.~Astron.~Soc.}}             
\def\prd{\aaref@jnl{Phys.~Rev.~D}}        
\def\prc{\aaref@jnl{Phys.~Rev.~C}}  
\def\prl{\aaref@jnl{Phys.~Rev.~Lett.}}    
\def\qjras{\aaref@jnl{QJRAS}}             
\def\skytel{\aaref@jnl{S\&T}}             
\def\ssr{\aaref@jnl{Space~Sci.~Rev.}}     
\def\zap{\aaref@jnl{ZAp}}                 
\def\nat{\aaref@jnl{Nature}}              
\def\aplett{\aaref@jnl{Astrophys.~Lett.}} 
\def\apspr{\aaref@jnl{Astrophys.~Space~Phys.~Res.}} 
\def\physrep{\aaref@jnl{Phys.~Rep.}}      
\def\physscr{\aaref@jnl{Phys.~Scr}}       
\def\commat{\aaref@jnl{Comm.~Math.~Phys.}}              
\def\science{\aaref@jnl{Science}}               
\def\cqg{\aaref@jnl{Classical Quant.~Grav.}}            
\def\jpcs{\aaref@jnl{JPCS}}                                     
\def\ijmpd{\aaref@jnl{Int.~J.~Mod.~Phys.~D}}                    
\def\grg{\aaref@jnl{Gen.~Relat.~Gravit.}}               
\def\rpp{\aaref@jnl{Rep.~Prog.~Phys.}}          
\def\npa{\aaref@jnl{Nucl.~Phys.~A}}        
\def\lrr{\aaref@jnl{Living Rev.~Rel.}}                   
\def\jcap{\aaref@jnl{J.~Cosmology Astropart.~Phys.}}    
\def\rmp{\aaref@jnl{Rev.~Mod.~Phys.}}   
\def\epjc{\aaref@jnl{Eur.~Phys.~J.~C}} 
\def\plb{\aaref@jnl{~Phy.~Lett.~B}} 
\def\mpla{\aaref@jnl{Mod.~Phy.~Lett.~A}} 
\def\arxiv{\aaref@jnl{arxiv.org}}

\allowdisplaybreaks[1]

\begin{document}
\color{black}       
\title{\bf Existence of non-exotic traversable wormholes in squared trace extended gravity theory}

\author{S. K. Tripathy \orcidlink{0000-0001-5154-2297}}
\email{tripathy\_sunil@rediffmail.com}
\affiliation{Department of Physics, Indira Gandhi Institute of Technology, Sarang, Dhenkanal, Odisha-759146, India.}

\author{D. Nayak}
\email{nayakdiptirekha283@gmail.com}
\affiliation{Department of Mathematics, Indira Gandhi Institute of Technology, Sarang, Dhenkanal, Odisha-759146, India.}

\author{B. Mishra\orcidlink{0000-0001-5527-3565}}
\email{bivu@hyderabad.bits-pilani.ac.in}
\affiliation{Department of Mathematics,
Birla Institute of Technology and Science-Pilani, Hyderabad Campus,
Hyderabad-500078, India.}

\author{D. Behera\orcidlink{0000-0002-5680-5794}}
\email{dipadolly@rediffmail.com}
\affiliation{Department of Physics, Indira Gandhi Institute of Technology, Sarang, Dhenkanal, Odisha-759146, India.}

\author{S. K. Sahu}
\email{jaga1282@gmail.com}
\affiliation{Department of Mathematics, Indira Gandhi Institute of Technology, Sarang, Dhenkanal, Odisha-759146, India.}


\begin{abstract}
An extended gravity theory is used to explore the possibility of non-exotic matter traversable wormholes. In the extended gravity theory, additional terms linear and quadratic in the trace of the energy momentum tensor are considered in the Einstein-Hilbert action. Obviously, such an addition leads to violation of the energy-momentum tensor. The model parameters are constrained from the structure of the field equations. Non-exotic matter wormholes tend to satisfy the null energy conditions. We use two different traversable wormhole geometries namely an exponential and a power law shape functions to model the wormholes. From a detailed analysis of the energy conditions, it is found that, the existence of non-exotic matter traversable wormholes is not obvious in the model considered and its possibility may depend on the choice of the wormhole geometry. Also, we found that, non-exotic wormholes are possible within the given squared trace extended gravity theory for a narrow range of the chosen equation of state parameter.
 
\end{abstract}

\maketitle
\textbf{Keywords}: Traversable wormholes; non-exotic matter; $f(R,T)$ gravity, Energy conditions

\section{Introduction} 

Modelling traversable wormholes in the frame work of extended gravity theories has gained a lot of research importance in recent times. In particular, the interest lies in the possibility of obtaining viable wormhole solutions with non-exotic matter content. Traversable wormholes are hypothetical tunnels connecting two distant parts of the same Universe or can also connect two separate manifolds \cite{Morris1988, Morris1988a}. They may look like a tube which becomes flat asymptotically on both the sides of the manifold. In general relativity (GR), traversable wormholes are possible with a violation of the null energy condition (NEC) as a consequence of satisfying the flare out condition \cite{Visser1995}. Such wormholes are required to contain exotic matter at least at the throat. Exotic matter wormholes may exists in both static \cite{Balakin2010, Anabalon2012} and dynamic  \cite{Bochicchio2010,Cataldo2011}  cases. The extended gravity theories proposed with an aim to explain the bizarre late time cosmic speed up issue, show the non-conservation of the energy-momentum tensor. The violation of energy-momentum conservation may provide an additional force  leading to the non-geodesic motion of the test particles. The structure of extended gravity theories with possible non-conservation of energy-momentum provides ways for modelling non-exotic traversable wormholes that satisfy NEC. The NEC for such exotic matter wormhole solutions may not be put as a strong constraint rather it appears as a weak condition on the associated geometry \cite{Capozziello2014, Hohmann2014}. 

With the advent of different cosmological observations leading to the prediction of late time cosmic acceleration, extended gravity theories have emerged as alternative routes to GR. One among these alternatives is the $f(R,T)$ gravity proposed by Harko et al. \cite{Harko2011} where $R$ and $T$ respectively denote the Ricci scalar and trace of energy-momentum tensor. The $f(R,T)$ theory has been successful in addressing different issues of cosmology and astrophysics. Of late, traversable wormhole models are also explored in this gravity. Some of the notable works are discussed here. The shape function of the wormhole can be obtained with the specification of equation of state parameter for the matter field \cite{Azizi2013}. The wormhole solutions can be obtained without the need of exotic matter \cite{Zubair2016} in few regions of the space time. The energy conditions of static wormhole has been shown and has been applied to study the physical and geometrical solutions \cite{Moraes2017}. Godani and Samanta \cite{Godani2019} have obtained the radius of the throat  of the wormhole and have minimised the presence of exotic matter.
The satisfaction of energy conditions in the matter that threads the wormhole throat has been shown in Ref. \cite{Banerjee2021}. The Casimir effect along with correction from the generalized uncertainty principle has been applied to model traversable wormholes in Ref. \cite{Tripathy2021}.  In the linear form of the functional $f(R,T)$, the non-exotic wormhole solutions can be realised \cite{Rosa2022} whereas the formation of specific static wormholes has been shown in Ref. \cite{Elizalde2018} . 

In the present work, we have considered an extended gravity theory where the action contains a quadratic term of the trace of the energy-momentum tensor in addition to the linear term and intending to search for the existence of viable non-exotic traversable wormhole solutions for two different geometries such as the exponential shape function and the power law shape function. The paper is organised as follows: in Sec.\ref{sec:II}, we present a basic theoretical framework of $f(R,T)$ gravity for the static spherically symmetric space time and its field equations. In Sec.\ref{sec:III}, we have presented the wormhole solutions and the energy conditions using two shape functions.  In Sec.\ref{sec:IV}, we have shown the exotic matter content of the tarversable wormhole and finally in Sec.\ref{sec:V}, the summary and conclusion are presented.

\section{Theoretical Framework}\label{sec:II}

We consider the action with a non-minimal coupling of matter-geometry within the setting up of a geometrically modified extended theory as 
\begin{equation} \label{eq:1}
S=\int d^4x\sqrt{-g}\left[ \frac{1}{2\kappa^2}f(R,T)+ \mathcal{L}_m \right],
\end{equation}
$\mathcal{L}_m$ being the matter Lagrangian and  $\kappa^2=8\pi$. This geometrically modified action has an arbitrary and well behaved function of $f(R,T)$ replacing the usual Ricci scalar $R$ in the Einstein-Hilbert action. Here $T$ is the trace  of the energy-momentum tensor. 

The functional $f(R,T)$ can be expressed as a sum of two separate functions and we may have $f(R,T)=f_1(R)+f_2(T)$ for a minimal matter-geometry coupling within the action.  

The modified field equation for the extended gravity theory may be obtained by a variation of the modified action with respect to the metric $g_{\mu\nu}$:

\begin{widetext}
\begin{equation} \label{eq:3}
R_{\mu\nu}-\frac{1}{2}f^{-1}_{1,R} (R)f_1(R)g_{\mu\nu}=f^{-1}_{1,R}(R)\left[\left(\nabla_{\mu} \nabla_{\nu}-g_{\mu\nu}\Box\right)f_{1,R}(R)+\left[\kappa^2 +f_{2,T}(T)\right]T_{\mu\nu}+\left[f_{2,T}(T)p+\frac{1}{2}f_2(T)\right]g_{\mu\nu}\right].
\end{equation}    
\end{widetext}    

In the above, we considered the shorthand notations for the partial derivatives with respect to the Ricci scalar and the trace of the energy-momentum tensor: $f_{1,R} (R)\equiv \frac{\partial f_1(R)}{\partial R}, f_{2,T} (T)\equiv \frac{\partial f_2(T)}{\partial T},  f^{-1}_{1,R} (R) \equiv \frac{1}{f_{1,R} (R)}$. Also, the matter Lagrangian is assumed to be the negative of the pressure of the cosmic fluid. The variation of the matter Lagrangian is associated with the energy-momentum tensor through the relation $T_{\mu\nu}=-\frac{2}{\sqrt{-g}}\frac{\delta\left(\sqrt{-g}\mathcal{L}_m\right)}{\delta g^{\mu\nu}}$.

We consider a simple choice $f_1(R)=R$ so that, the modified theory provides GR like field equations 
\begin{equation}\label{eq:6}
G_{\mu\nu}= \left[\kappa^2 +f_{2,T}(T)\right]T_{\mu\nu}+\left[f_{2,T}(T)p+\frac{1}{2}f_2(T)\right]g_{\mu\nu}.
\end{equation}
The above equation may be written as
\begin{equation}\label{eq:7}
G_{\mu\nu}= \kappa^2_{T}\left[T_{\mu\nu}+ T^{int}_{\mu\nu}\right],
\end{equation}
where $\kappa^2_{T}= \kappa^2 +f_{2,T}(T)$ , $G_{\mu\nu}$ is the Einstein tensor and we have an interactive term
\begin{equation}\label{eq:8}
T^{int}_{\mu\nu}=\left[ \frac{f_{2,T}(T)p+\frac{1}{2}f_2(T)}{\kappa^2 +f_{2,T}(T)}\right]g_{\mu\nu}.
\end{equation}
This interactive term  may be due to the presence of imperfect fluids  or certain quantum effects evolved in the field equations because of the geometry modification of the action. The existence of the interactive term provides us an idea that, the non-minimal matter-geometry coupling in the action generates an additional fictitious matter field which may shoulder the responsibility to explain the late time cosmic speed up phenomena.

It should be remarked here that, a linear choice of the functional $f_2(T)$ leads to a constant redefined Einstein constant $\kappa^2_T$. However, in the present work, we consider a quadratic functional $f_2(T)=\lambda_1 T +\lambda_2 T^2$ so that, the redefined Einstein constant becomes time dependent. $\lambda_1$ and $\lambda_2$ are the model parameters and to be fixed from the specific consideration of the wormhole solutions. For such a choice, the interactive terms becomes
\begin{equation}
T^{int}_{\mu\nu}=\frac{1}{\kappa^2_{T}}\left[\lambda_1\left(p+\frac{T}{2}\right)+\lambda_2 T\left(2p+\frac{T}{2}\right)\right]g_{\mu\nu}.
\end{equation}

The construction of traversable wormholes involve the assumption of an anisotropic fluid content represented by the energy-momentum tensor
\begin{equation}
T_{\mu}^{\nu}= diag (\rho, -p_r, -p_t, -p_t),
\end{equation}
where $p_r$ and $p_t$ are respectively the radial and tangential pressure components. $\rho$ is the energy density of the matter content. The total pressure for the cosmic fluid becomes $p=\frac{1}{3}(p_r+2p_t)$ and the trace of the energy-momentum tensor is $T=\rho-3p$. Spherical symmetry preservation and time independent nature of the solutions require that the energy density, the radial and tangential component of the fluid pressure be functions of only radial coordinates.

We consider static and spherically symmetric wormholes described through the metric
\begin{equation}
ds^2=-e^{\Phi(r)}dt^2+\left(1-\frac{b(r)}{r}\right)^{-1}dr^2+r^2d\Omega^2,
\end{equation}
where $\Phi(r)$ and $b(r)$ are respectively the redshift function and the shape function depending only on the radial coordinates. $d\Omega^2=d\theta^2+\sin^2\theta d\phi^2$ is the surface element. For wormholes with zero tidal force, the redshift function may be considered as constant or zero. Also, it is worth mentioning here that, wormhole solutions with $\Phi\neq 0$ and $\Phi=0$ do not have much difference. In view of this, we consider wormhole solutions with constant redshift function. The field equations for the traversable wormhole metric become
\begin{widetext}
\begin{eqnarray}
\frac{b^{\prime}(r)}{r^2} &=& 8\pi\rho+\frac{\lambda_1}{6}\left[9\rho-(p_r+2p_t)\right]+\frac{\lambda_2}{6}\left[(\rho-p_r-2p_t)(15\rho+p_r+2p_t)\right],\label{eq:9}\\
\frac{b(r)}{r^3} &=& -8\pi p_r+\frac{\lambda_1}{6}\left[3\rho-(7p_r+2p_t)\right]+\frac{\lambda_2}{6}\left[(\rho-p_r-2p_t)(3\rho-11p_r+2p_t)\right],\label{eq:10}\\
\frac{b^{\prime}(r)}{r^2}-\frac{b(r)}{r^3} &=& -16\pi p_t+\frac{\lambda_1}{3}\left[3\rho-(p_r+8p_t)\right]+\frac{\lambda_2}{3}\left[(\rho-p_r-2p_t)(3\rho+p_r-10p_t)\right].\label{eq:11}
\end{eqnarray}    
\end{widetext}

Traversable wormhole solutions may be obtained with the fluid content satisfying the barotropic equations of state i.e. $p_r=\omega_r\rho$ and $p_t=\omega_t\rho$ where $\omega_r$ and $\omega_t$ are the equation of state (EoS) parameters. Depending upon the requirement one may chose these EoS parameters arbitrarily or may fix from certain physical basis. The field equations \eqref{eq:9}-\eqref{eq:11} may be written as
\begin{eqnarray}
\frac{b^{\prime}(r)}{r^2} &=& \xi_1\rho+\gamma~\xi_2\rho^2,\label{eq:12}\\
\frac{b(r)}{r^3} &=& \xi_3\rho+\gamma~\xi_4\rho^2,\label{eq:13}\\
\frac{b^{\prime}(r)}{r^2}-\frac{b(r)}{r^3} &=& \xi_5\rho+\gamma~\xi_6\rho^2,\label{eq:14}
\end{eqnarray}
where 
\begin{eqnarray}
\xi_1 &=&  \frac{\lambda_1}{6}\left(9-\omega_r-2\omega_t\right)+8\pi , ~\xi_2 = \omega_r+2\omega_t+15, \nonumber\\
\xi_3 &=& \frac{\lambda_1}{6}\left(3-7\omega_r-2\omega_t\right)-8\pi\omega_r, ~\xi_4 = 3-11\omega_r+2\omega_t, \nonumber \\
\xi_5 &=& \frac{\lambda_1}{3} \left(3-\omega_r-8\omega_t\right)-16\pi\omega_t, ~\xi_6 = 2\left(3+\omega_r-10\omega_t\right),\nonumber\\
\gamma &=& \frac{\lambda_2}{6} \left(1-\omega_r-2\omega_t\right).  \label{eq:15}
\end{eqnarray}

The energy density for the traversable wormhole may be obtained from Eqs.\eqref{eq:12} and \eqref{eq:13} as
\begin{equation}
\rho=\left[\frac{\xi_4 r\frac{b^{\prime}(r)}{b(r)}-\xi_2}{\xi_1\xi_4-\xi_2\xi_3}\right]\frac{b(r)}{r^3}.\label{eq:16}
\end{equation}

The energy density obviously depends on the model parameters, the EoS parameters and the choice of the shape function for the traversable wormhole. The extended gravity theory we considered here contains two arbitrary parameters $\lambda_1$ and $\lambda_2$. 

From Eqs. \eqref{eq:12}- \eqref{eq:14}, we obtain certain restrictions on the model parameters $\lambda_1, \omega_r$ and $\omega_t$ through the relations
\begin{eqnarray}
\xi_1-\xi_3 &=& \xi_5,\label{eq:17}\\
\gamma~(\xi_2-\xi_4) &=& \gamma~\xi_6.\label{eq:18}
\end{eqnarray}

Imposition of the condition in Eq.\eqref{eq:17}, we get a relationship between the radial and the tangential pressure as
\begin{equation}
\omega_r=-(2\omega_t+\chi),\label{eq:19}
\end{equation}
where $\chi=\frac{6\pi}{\lambda_1+6\pi}$. From the second condition in Eq.\eqref{eq:18} along with a non-vanishing trace of the matter field (i.e. $\gamma\neq 0$), we obtain a constraint on the model parameter $\lambda_1$ as $\lambda_1=4\pi$. However, for a vanishing trace of the matter field ($\gamma=0$), we get the constraint $\lambda_1=-12\pi$. Throughout this work, we use $\lambda_1=4\pi$ so that the trace of the matter content of the traversable wormhole should not vanish. One should note that, the numerical values of these parameters are very much constrained from the model itself which clearly restricts the arbitrarily chosen values for  these parameters.

It is interesting to note here that, the energy density of the traversable wormhole model does not depend on the choice of the modified gravity parameter $\lambda_2$ and the other parameter $\lambda_1$ may be fixed from the restrictions on the field equations. Since the energy density does not depend on $\lambda_2$, the coupling constant of the squared trace term in the action, all other relevant quantities such as the pressure and energy conditions may not depend on $\lambda_2$. However, this does not necessarily mean that, the functional $f(R,T)$ may be reduced to the linear term in $T$ which leads to the unacceptable condition $\gamma=0$. 

\section{Wormhole solutions and Energy Conditions}\label{sec:III}

In order to have some physically viable and reasonable models of traversable wormholes and to investigate the possibility of non-exotic wormholes within the squared trace extended gravity theory, we consider two different shape functions namely an exponential shape function $b(r)=re^{-2(r-r_0)}$ and a power law shape function $b(r)=\sqrt{r_0r}$, where $r_0$ is the throat radius of the wormhole. One should note that, these shape functions should satisfy some of the conditions such that it should reproduce the throat radius at the throat i.e $b(r_0)=r_0$, the slope of the shape function at the throat should be less than unity, $b^{\prime}(r_0)\leq 1$. Other necessary conditions, the wormhole shape function should satisfy include the metric condition $1-\frac{b(r)}{r}\geq 0$ and the flare out condition $\frac{b(r)-rb^{\prime}(r)}{b^2(r)}>0$.

In the classical GR, violation of the energy conditions is an essential feature for the wormhole geometry. Usually, for exotic matter content of the wormhole, the null energy conditions $\rho+p_r\geq 0$ (NEC1) and $\rho+p_t \geq 0$ (NEC2)  including the strong energy conditions $\rho+p_r+2p_t \geq 0$ (SEC) are violated. However, the scenario is changed in the geometrically modified gravity theories where, the congruence of the time like and the space like curves in Raychaudhuri equation involving the term $R_{\mu\nu}u^{\mu}u^{\nu}$, where $u^{\mu}$ is the space like null vector, is not very obvious \cite{Mishra2021,Mishra2022}. This situation alters the so called violation of the energy condition of wormhole geometry in extended gravity theories and may lead to the existence of non-exotic traversable wormholes.

\subsection {Case-I}
As a first example, we consider an exponential shape function described through $b(r)=re^{-2(r-r_0)}$. The slope of this shape function becomes $b^{\prime}(r)=(1-2r)\frac{b(r)}{r}$ and we have $\frac{b^{\prime}(r)}{b(r)}=\frac{1}{r}-2$. We plot the shape function and its derivatives in FIG.1. It is shown in the figure that, the shape function satisfies all the requirements to be acceptable as a possible wormhole geometry.

\begin{figure}[H]
\centering
\includegraphics[width=0.5\textwidth]{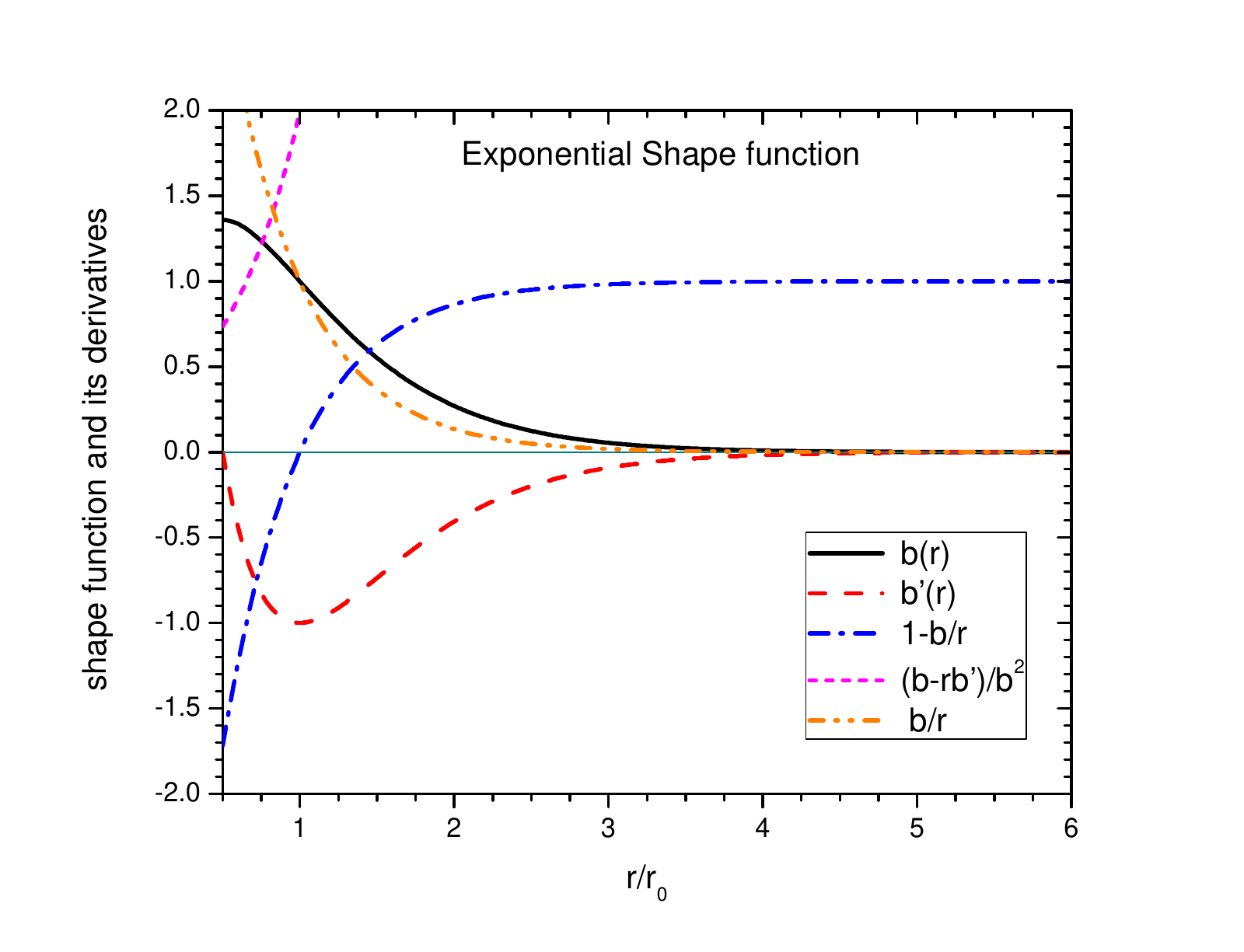}
\caption{Exponential shape function and its behaviour.}\label{FIG.1}
\end{figure}

With this shape function, the energy density, radial pressure and the tangential pressure become
\begin{widetext}
\begin{eqnarray}
\rho &=& \left[\frac{-2r(3-11\omega_r+2\omega_t)-12(\omega_r+1)}{(2\omega_r\omega_t+2\omega_t+\omega_r^2+3)(8\pi+3\lambda_1)-12\lambda_1+32\pi\omega_r}\right]r^{-2}e^{-2(r-r_0)},\\
p_r &=& \left[\frac{2r(3-11\omega_r+2\omega_t)+12(\omega_r+1)}{(2\omega_r\omega_t+2\omega_t+\omega_r^2+3)(8\pi+3\lambda_1)-12\lambda_1+32\pi\omega_r}\right](2\omega_t+\chi)r^{-2}e^{-2(r-r_0)},\\
p_t &=& \left[\frac{-2r(3-11\omega_r+2\omega_t)-12(\omega_r+1)}{(2\omega_r\omega_t+2\omega_t+\omega_r^2+3)(8\pi+3\lambda_1)-12\lambda_1+32\pi\omega_r}\right]\omega_t~r^{-2}e^{-2(r-r_0)}.
\end{eqnarray}    
\end{widetext}

The pressure anisotropy for this wormhole geometry will be
\begin{widetext}
\begin{equation}
\triangle p = p_r-p_t = \left[\frac{2r(3-11\omega_r+2\omega_t)+12(\omega_r+1)}{(2\omega_r\omega_t+2\omega_t+\omega_r^2+3)(8\pi+3\lambda_1)-12\lambda_1+32\pi\omega_r}\right](3\omega_t+\chi)r^{-2}e^{-2(r-r_0)}.
\end{equation}    
\end{widetext}

We wish to investigate the existence of non-exotic traversable wormholes within the extended gravity theory. Another important fact is the presumption of the tangential and/or the radial pressure of the matter content. Two possibilities have been explored in the present work. One is the assumption of an exotic fluid pressure with a negative value ($\omega_t=-1.03$) and the other corresponds to the assumption of a realistic fluid with positive pressure ($\omega_t=0.95$). The corresponding value for $\omega_r$ becomes $-2.66$ and $-2.3$ respectively. In FIG.2, we show the evolution of the energy density, radial pressure, tangential pressure and the pressure anisotropy for  $\omega_t=-1.03$ (Left panel) and for $\omega=0.95$ (Right panel). For the figure with $\omega_t=-1.03$, the radial pressure and the energy density are positive upto $r=0.97r_0$ and then they become negative. On the other hand, the tangential pressure is negative for $r < 0.97r_0$ and beyond this radial distance it becomes positive. However, all of them have vanishing small values at large radial distance. The pressure anisotropy within the wormhole geometry evolves from a positive value, becomes negative after $r=0.97r_0$ and washes out at large radial distance. For the figure with $\omega=0.95$ (Right panel of FIG.2), the energy density and the tangential pressure are negative throughout whereas the radial pressure and the pressure anisotropy remain in the positive domain. Also, the pressure anisotropy vanishes for large radial distance. 

\begin{figure*}
\centering
\minipage{0.50\textwidth}
\includegraphics[width=\textwidth]{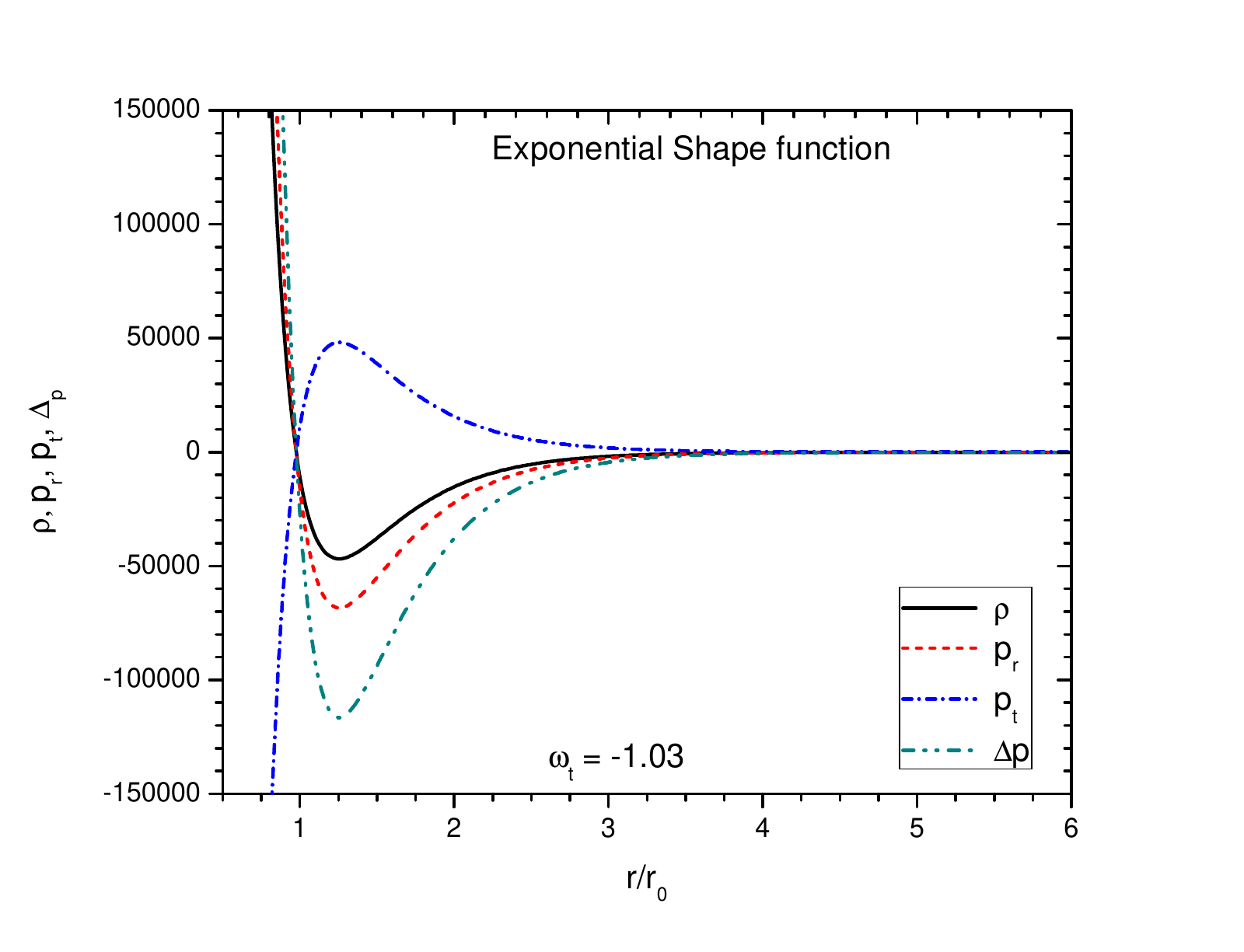}
\endminipage\hfill
\minipage{0.50\textwidth}
\includegraphics[width=\textwidth]{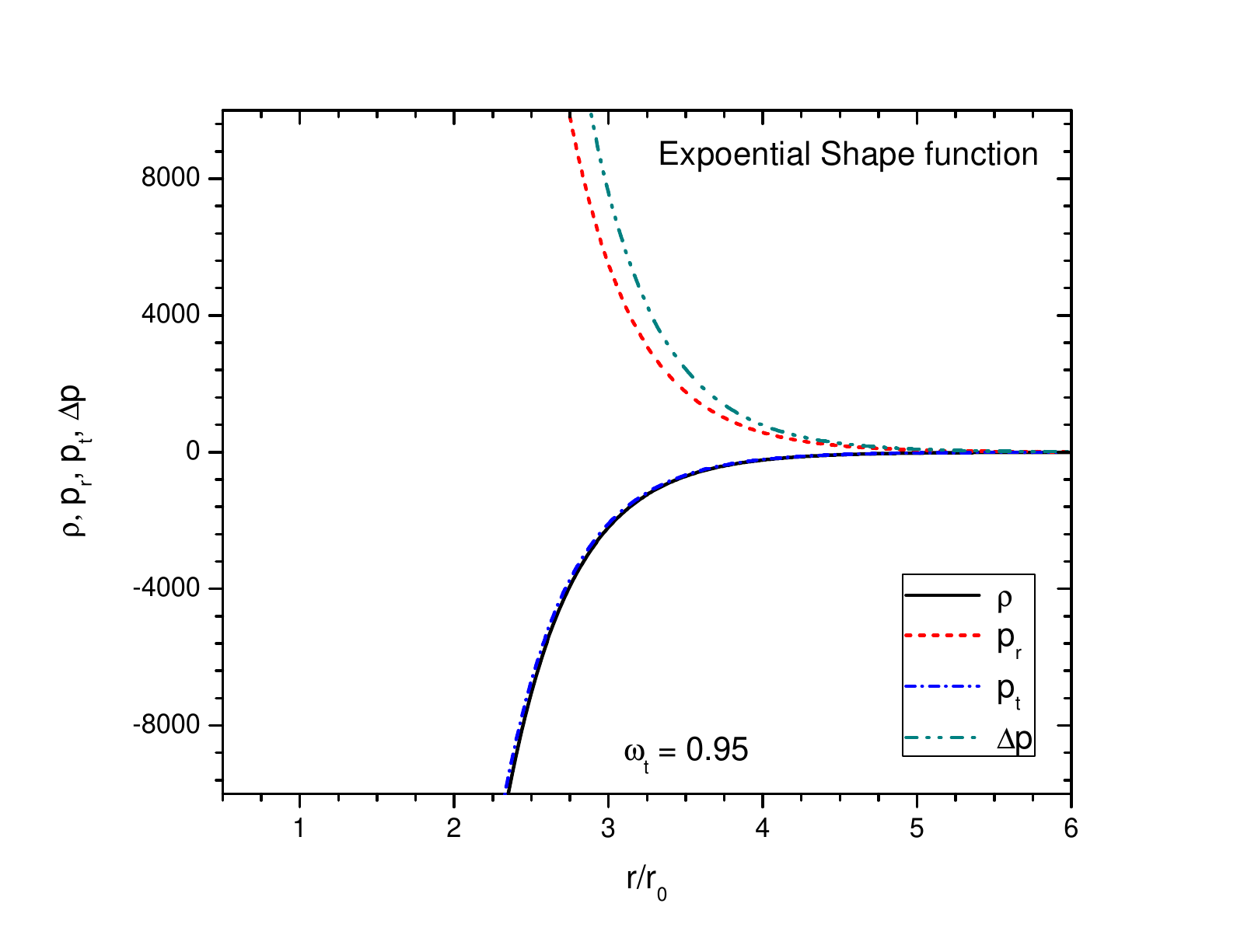}
\endminipage\hfill
\caption{Behaviour of energy density, radial pressure, tangential pressure and the pressure anisotropy assuming an exponential shape function. For  $\omega_t=-1.03$ (Left Panel) and for  $\omega_t=0.95$ (Right Panel).}\label{FIG.2}
\end{figure*}

The energy conditions for the traversable wormholes with an exponential shape functions may be obtained as
\begin{widetext}
\begin{eqnarray}
\rho+p_r &=& \frac{2\omega_t+\chi-1}{r^{2}e^{2(r-r_0)}}\left[\frac{2r(3-11\omega_r+2\omega_t)+12(\omega_r+1)}{(2\omega_r\omega_t+2\omega_t+\omega_r^2+3)(8\pi+3\lambda_1)-12\lambda_1+32\pi\omega_r}\right],\\
\rho+p_t &=& \frac{1+\omega_t}{r^{2}e^{2(r-r_0)}}\left[\frac{-2r(3-11\omega_r+2\omega_t)-12(\omega_r+1)}{(2\omega_r\omega_t+2\omega_t+\omega_r^2+3)(8\pi+3\lambda_1)-12\lambda_1+32\pi\omega_r}\right],\\
\rho+p_r+2p_t &=& \frac{\chi-1}{r^{2}e^{2(r-r_0)}}\left[\frac{2r(3-11\omega_r+2\omega_t)+12(\omega_r+1)}{(2\omega_r\omega_t+2\omega_t+\omega_r^2+3)(8\pi+3\lambda_1)-12\lambda_1+32\pi\omega_r}\right].
\end{eqnarray}    
\end{widetext}

Other relevant energy conditions may be
\begin{widetext}
\begin{eqnarray}
\rho-p_r &=& \frac{1+2\omega_t+\chi}{r^{2}e^{2(r-r_0)}}\left[\frac{-2r(3-11\omega_r+2\omega_t)-12(\omega_r+1)}{8\pi(4\omega_r+2\omega_r\omega_t+2\omega_t+\omega_r^2+3)+3\lambda_1(\omega_r^2+2\omega_r\omega_t+2\omega_t-1)}\right], \\
\rho-p_t &=& \frac{\omega_t-1}{r^{2}e^{2(r-r_0)}}\left[\frac{2r(3-11\omega_r+2\omega_t)+12(\omega_r+1)}{8\pi(4\omega_r+2\omega_r\omega_t+2\omega_t+\omega_r^2+3)+3\lambda_1(\omega_r^2+2\omega_r\omega_t+2\omega_t-1)}\right].
\end{eqnarray}    
\end{widetext}

The energy conditions for the wormhole with exponential shape function are shown in FIG.3. For this class of wormhole geometry, the SEC is satisfied within $r=0.97r_0$ for $\omega_t=-1.03$ and is violated for the remaining region. For the case with $\omega_t=0.95$, SEC is violated through out the radial domain. With $\omega_t=0.95$, while the wormhole matter content satisfies the NEC1, it violates NEC2. On the other hand, for $\omega_t=-1.03$, we have a varied picture regarding the violation of NEC. Within $r=0.97r_0$, the NEC1 is satisfied and NEC2 is violated. However beyond $r=0.97r_0$, the situation reverses i.e, NEC1 is violated and NEC2 is satisfied. 

\begin{figure*}
\centering
\minipage{0.50\textwidth}
\includegraphics[width=\textwidth]{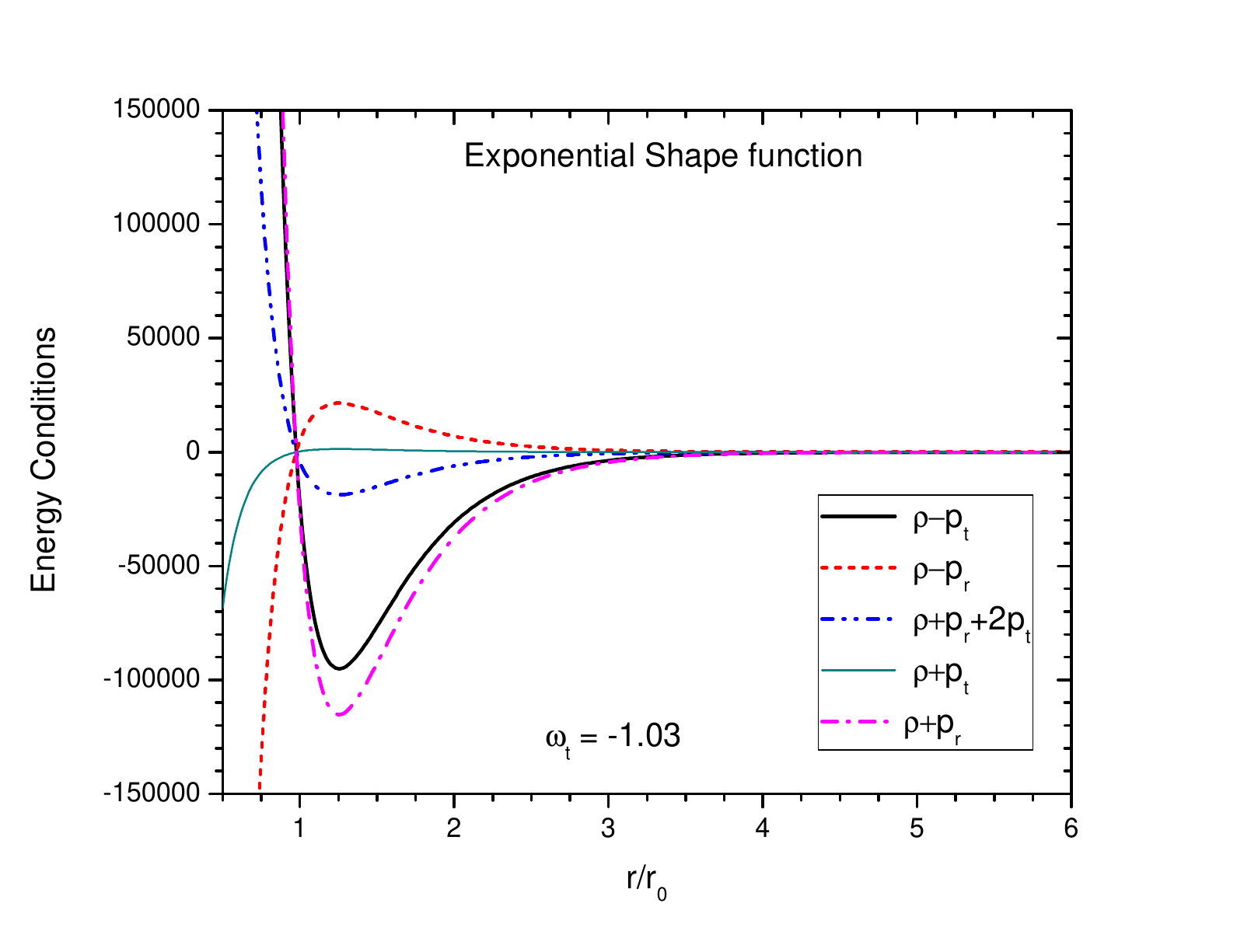}
\endminipage\hfill
\minipage{0.50\textwidth}
\includegraphics[width=\textwidth]{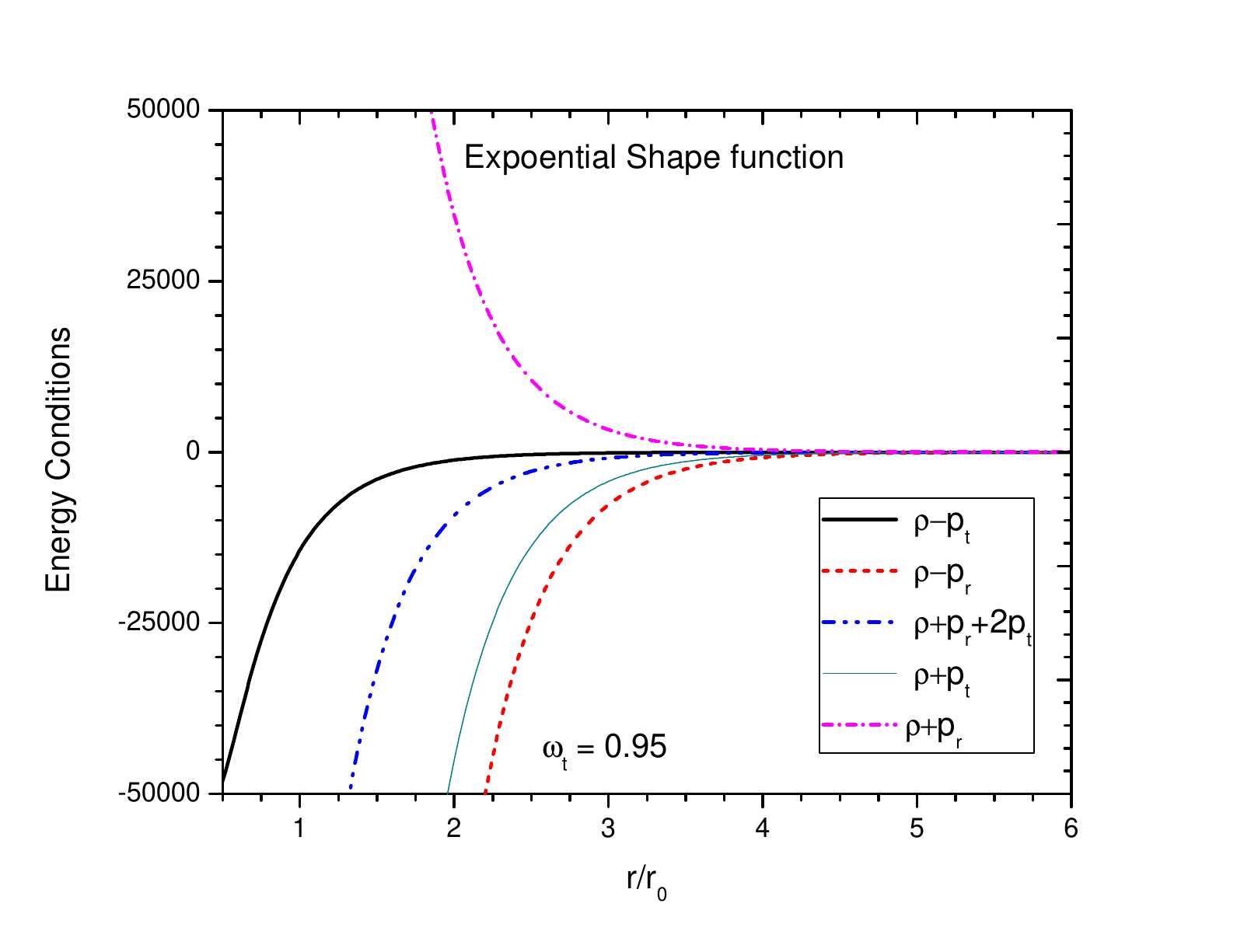}
\endminipage\hfill
\caption{Energy Conditions for the traversable wormholes with exponential shape function for $\omega=-1.03$ (Left Panel) and for $\omega=0.95 $ (Right Panel).}\label{FIG.3}
\end{figure*}

In order to check whether, it is possible to obtain traversable wormholes with non-exotic matter content for an exponential shape function, in FIG.4, we plot $\rho+p=\rho+\frac{1}{3}(p_r+2p_t)$ to get an idea about satisfaction of the average null energy condition i.e. $\rho+p\geq 0$ (ANEC). For both the choices of the EoS parameter, the ANEC is violated after $r=0.97r_0$. However, the ANEC is satisfied for the $\omega_t=-1.03$ case prior to $r=0.97r_0$.
\begin{figure}[H]
\centering
\includegraphics[width=0.5\textwidth]{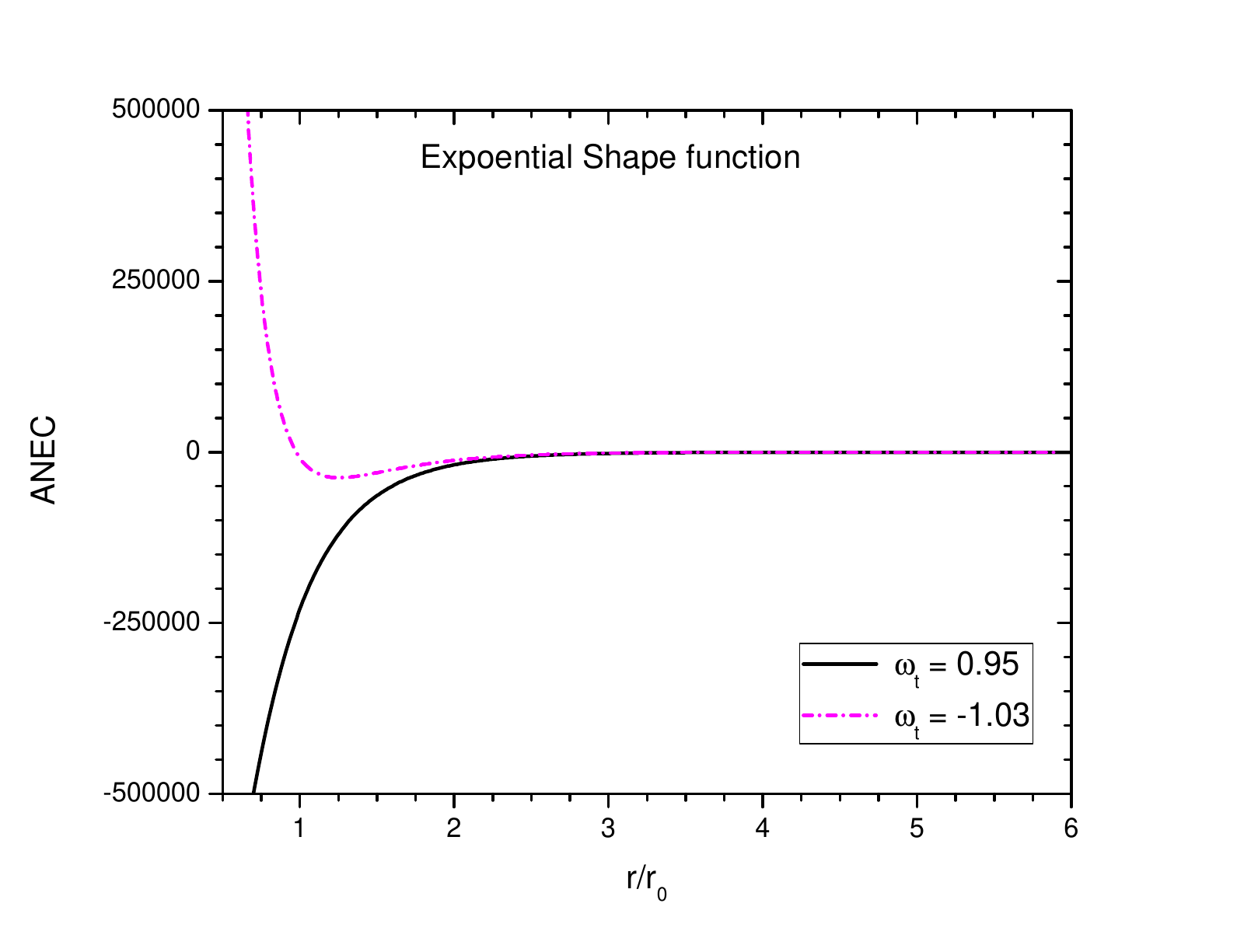}
\caption{ANEC for the exponential shape function.}\label{FIG.4}
\end{figure}
We have investigated with different set of EoS parameters in the range $-1.1\leq \omega_t \leq 1$ with an aim to get both the NEC1 and NEC2 to be satisfied along with the satisfaction of the ANEC. The results of our investigation are presented in TABLE-I. One may observe that, for positive values of $\omega_t$ within the range specified, it is not possible to get traversable wormholes with simultaneous satisfaction of NEC1 and NEC2. On the other hand, in the negative domain of $\omega_t$, we do not find the possibility of non-exotic wormholes upto $\omega_t=-0.36$. However, in the range $-0.37\leq \omega_t \leq -0.44$, there is possibility of the existence of non-exotic wormholes with exponential shape function within the present squared trace gravity theory. In the range $-0.45\leq \omega_t \leq 1$, the NEC1, NEC2 and ANEC are satisfied upto certain radial distance as mentioned in TABLE-I. It is observed that, the radial distance upto which there is satisfaction of these energy conditions decreases as we decrease the value of $\omega_t$ from $-0.45$ to $-1.0$. Below the choice $\omega_t=-1.0$, it is not possible to have non-exotic wormholes as simultaneous satisfaction of NEC1 and NEC2 is not possible. 

\begin{table}
\caption{The range of parameter for $\omega_t$ for which NEC1, NEC2 and ANEC are satisfied or violated for the exponential shape function. A violation of the energy condition is denoted through ($\times$) and their satisfaction by ($\surd$).}
\begin{tabular}{c|c|c|c|c}
\hline
range of $\omega_t$ & NEC1      & NEC2      & ANEC       & \begin{tabular}{@{}c@{}}satisfied
     \\upto  ($r/r_0$) 
\end{tabular} \\
\hline
1					& $\times$	& $\surd$	& $\surd$ & \\
0.99 - 0.3			& $\surd$	& $\times$	&   $\times$	&\\
0.1 - $-0.36$		& $\times$	& $\times$	&	$\times$	& \\
$-0.37$ - $-0.44$	& $\surd$	&  $\surd$	& $\surd$ &\\
$-0.5$				& $\surd$ 	& $\surd$	&$\surd$ & 3.5\\
$-0.7$				& $\surd$ 	& $\surd$	&$\surd$ & 1.5\\
$-0.8$				& $\surd$ 	& $\surd$	&$\surd$ & 1.25\\
$-0.9$				& $\surd$ 	& $\surd$	&$\surd$ & 1.09\\
$-1$				& $\surd$ 	& $\surd$	&$\surd$ & 1.0\\
\hline
\end{tabular}
\end{table}

\subsection {Case-II}
Here we consider a power law shape function for the wormhole geometry described through $b(r)=\sqrt{r_0r}$ whose slope is $b^{\prime}(r)=\frac{1}{2}\sqrt{\frac{r_0}{r}}$ and  $\frac{b^{\prime}(r)}{b(r)}=\frac{1}{2r}$. This shape function is the same geometry assumed in Ref. \cite{Mishra2021}. We plot the shape function in FIG.5 and one may find that the shape function satisfies the basic requirements.

\begin{figure}[ht!]
\centering
\includegraphics[width=0.5\textwidth]{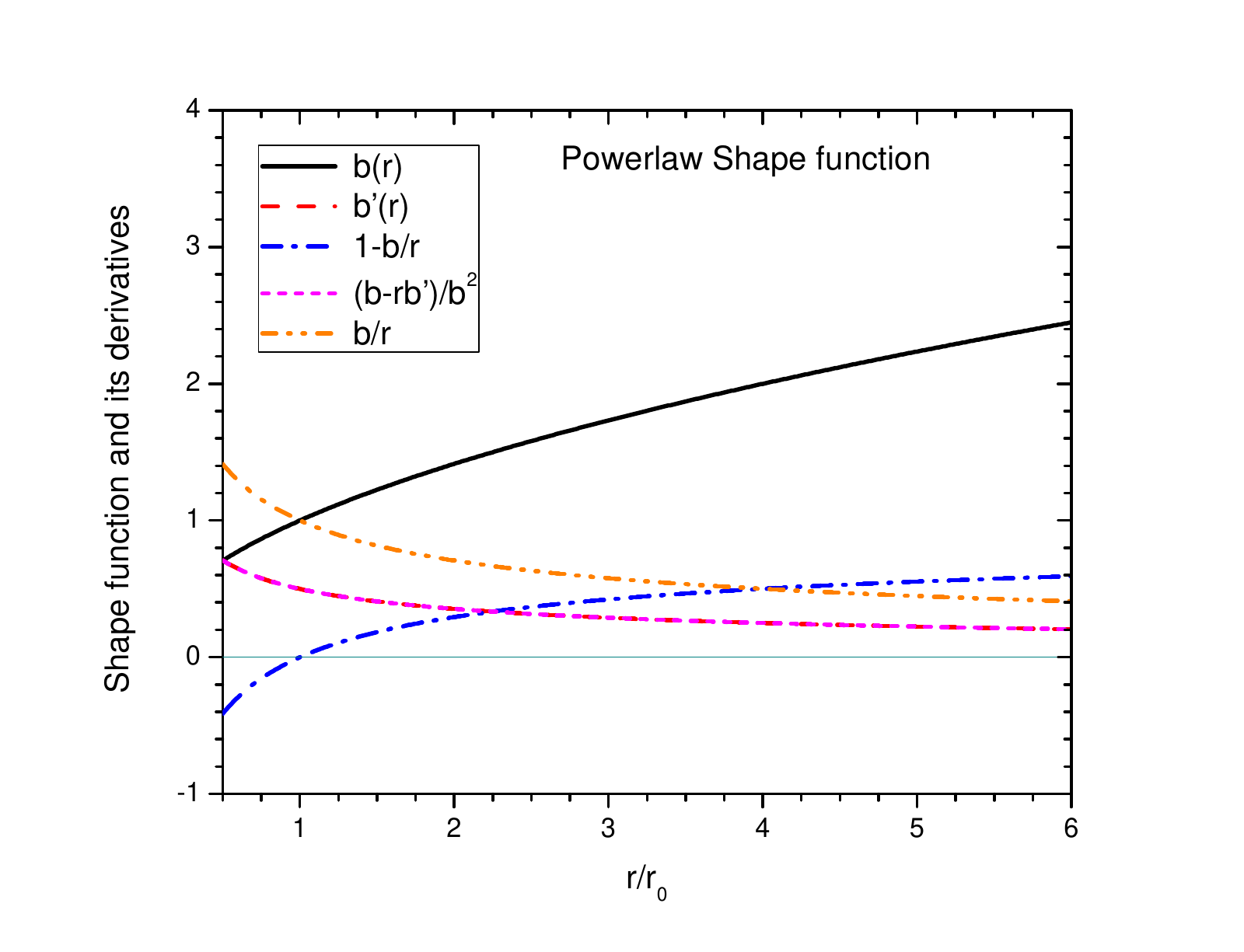}
\caption{Power law shape function and its behaviour.}\label{FIG.5}
\end{figure}

The energy density, radial pressure and the tangential pressure for this shape function are obtained as
\begin{widetext}
\begin{eqnarray}
\rho &=& \frac{1}{2}\left[\frac{-27-13\omega_r-2\omega_t}{(2\omega_r\omega_t+2\omega_t+\omega_r^2+3)(8\pi+3\lambda_1)-12\lambda_1+32\pi\omega_r}\right]\sqrt{\frac{r_0}{r^5}},\\
p_r &=& \frac{1}{2}\left[\frac{27+13\omega_r+2\omega_t}{(2\omega_r\omega_t+2\omega_t+\omega_r^2+3)(8\pi+3\lambda_1)-12\lambda_1+32\pi\omega_r}\right](2\omega_t+\chi)\sqrt{\frac{r_0}{r^5}},\\
p_t &=& \frac{1}{2}\left[\frac{-27-13\omega_r-2\omega_t}{(2\omega_r\omega_t+2\omega_t+\omega_r^2+3)(8\pi+3\lambda_1)-12\lambda_1+32\pi\omega_r}\right]\omega_t\sqrt{\frac{r_0}{r^5}}.
\end{eqnarray}    
\end{widetext}

The pressure anisotropy for this wormhole geometry becomes
\begin{widetext}
\begin{equation}
\triangle p = \frac{1}{2}\left[\frac{27+13\omega_r+2\omega_t}{(2\omega_r\omega_t+2\omega_t+\omega_r^2+3)(8\pi+3\lambda_1)-12\lambda_1+32\pi\omega_r}\right](3\omega_t+\chi)\sqrt{\frac{r_0}{r^5}}.
\end{equation}    
\end{widetext}

The evolution of the energy density, radial pressure, tangential pressure and the pressure anisotropy for the chosen two values of the EoS parameter namely $\omega_t=-1.03$ (left panel) and for $\omega=0.95$ (right panel) are shown in FIG.6. For  both the cases, the energy density remains positive  and evolve to null values for large radial distance. While the tangential pressure is positive for $\omega=0.95$, it is negative for $\omega_t=-1.03$. The radial pressure and the pressure anisotropy show an opposite behaviour to the tangential pressure. They are positive for $\omega_t=-1.03$ and negative for $\omega=0.95$. However for large values of $r$, all of these quantities become null.

\begin{figure*}
\centering
\minipage{0.50\textwidth}
\includegraphics[width=\textwidth]{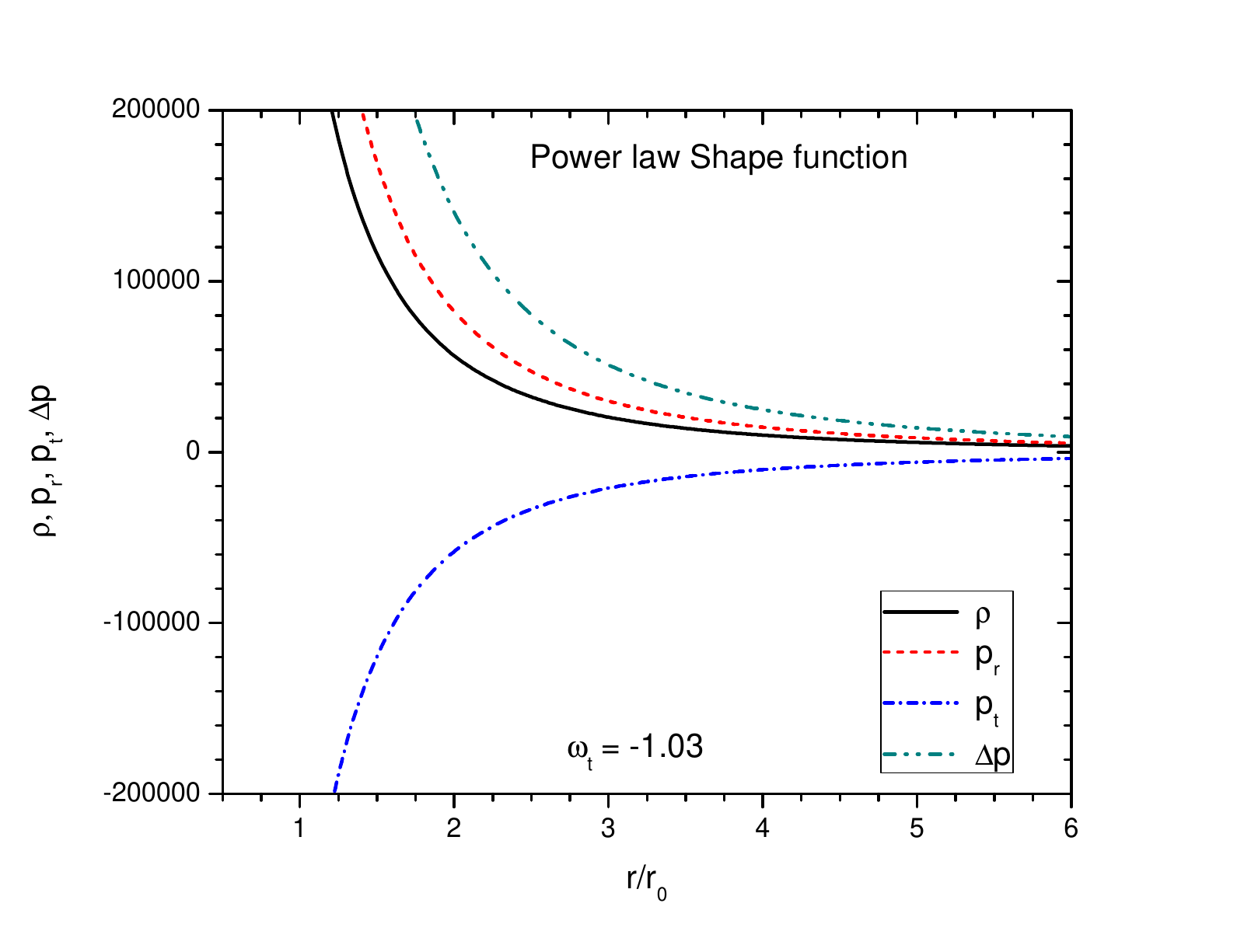}
\endminipage\hfill
\minipage{0.50\textwidth}
\includegraphics[width=\textwidth]{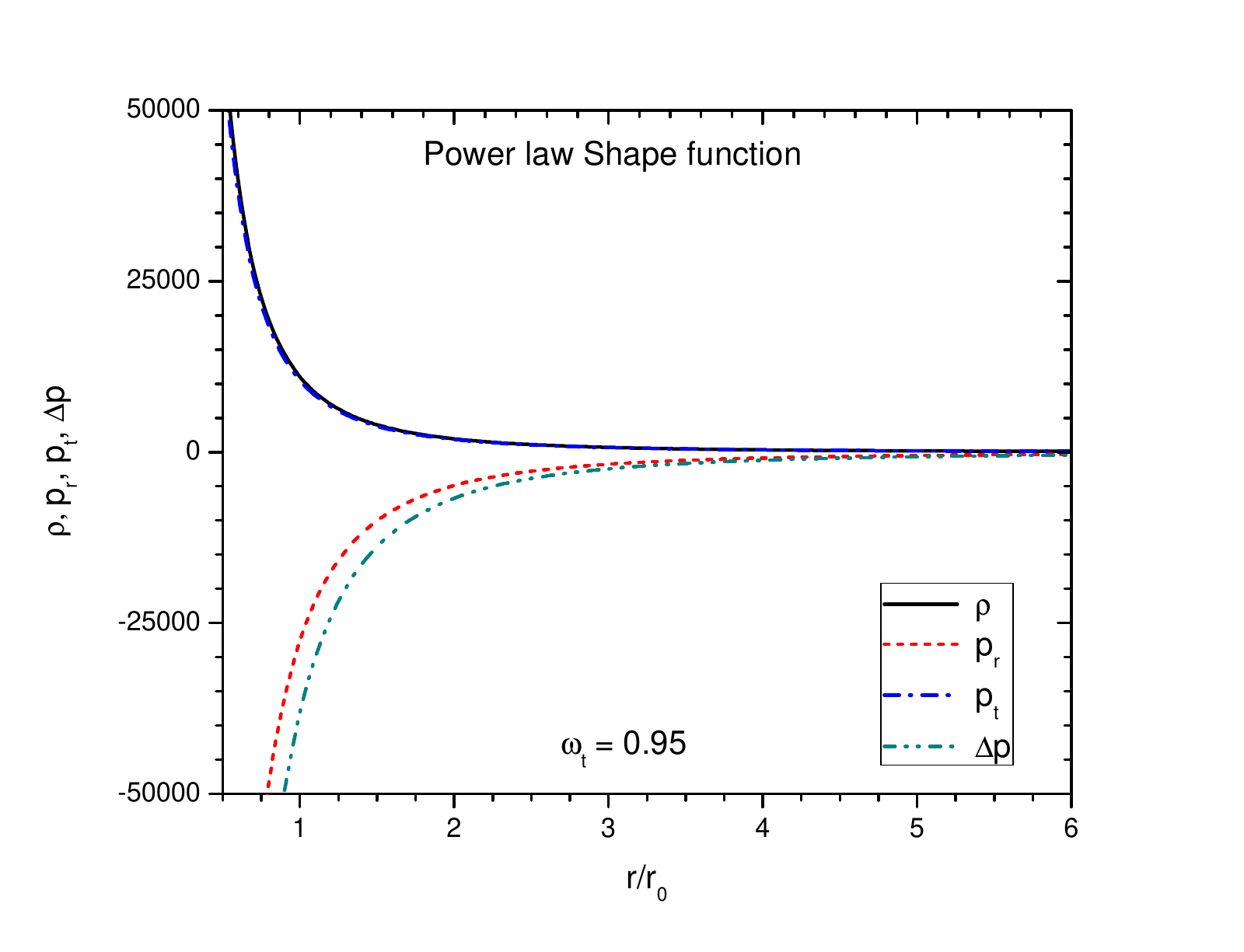}
\endminipage\hfill
\caption{Behaviour of energy density, radial pressure, tangential pressure and the pressure anisotropy assuming a power law shape function. For  $\omega_t=-1.03$ (Left Panel) and for  $\omega_t=0.95$ (Right Panel).}\label{FIG.6}
\end{figure*}

It is straight forward to obtain the behaviour of the energy conditions for the power law shape functions from the expressions of the energy density, tangential and radial pressure obtained above:
\begin{widetext}
\begin{eqnarray}
\rho+p_r &=& \frac{2\omega_t+\chi-1}{2}\left[\frac{27+13\omega_r+2\omega_t}{(2\omega_r\omega_t+2\omega_t+\omega_r^2+3)(8\pi+3\lambda_1)-12\lambda_1+32\pi\omega_r}\right]\sqrt{\frac{r_0}{r^5}},\\
\rho+p_t &=& \frac{1+\omega_t}{2}\left[\frac{-27-13\omega_r-2\omega_t}{(2\omega_r\omega_t+2\omega_t+\omega_r^2+3)(8\pi+3\lambda_1)-12\lambda_1+32\pi\omega_r}\right]\sqrt{\frac{r_0}{r^5}},\\
\rho+p_r+2p_t &=& (\chi-1)\frac{1}{2}\left[\frac{27+13\omega_r+2\omega_t}{(2\omega_r\omega_t+2\omega_t+\omega_r^2+3)(8\pi+3\lambda_1)-12\lambda_1+32\pi\omega_r}\right]\sqrt{\frac{r_0}{r^5}},\\
\rho-p_r &=&\frac{1+2\omega_t+\chi}{2}\left[\frac{-27-13\omega_r-2\omega_t}{(2\omega_r\omega_t+2\omega_t+\omega_r^2+3)(8\pi+3\lambda_1)-12\lambda_1+32\pi\omega_r}\right]\sqrt{\frac{r_0}{r^5}}, \\
\rho-p_t &=& \frac{\omega_t-1}{2}\left[\frac{27+13\omega_r+2\omega_t}{(2\omega_r\omega_t+2\omega_t+\omega_r^2+3)(8\pi+3\lambda_1)-12\lambda_1+32\pi\omega_r}\right]\sqrt{\frac{r_0}{r^5}}.
\end{eqnarray}
    
\end{widetext}

We plot the energy conditions for the matter content of the wormhole described through the power law shape function in FIG.7. It is found that, while the NEC1 is satisfied, the NEC2 is violated for the whole range of radial coordinate for the wormhole geometry with $\omega_t=-1.03$ and the reverse is the case for $\omega_t=0.95$. It is interesting to note that, the SEC is satisfied in both the cases. 
\begin{figure*}
\centering
\minipage{0.50\textwidth}
\includegraphics[width=\textwidth]{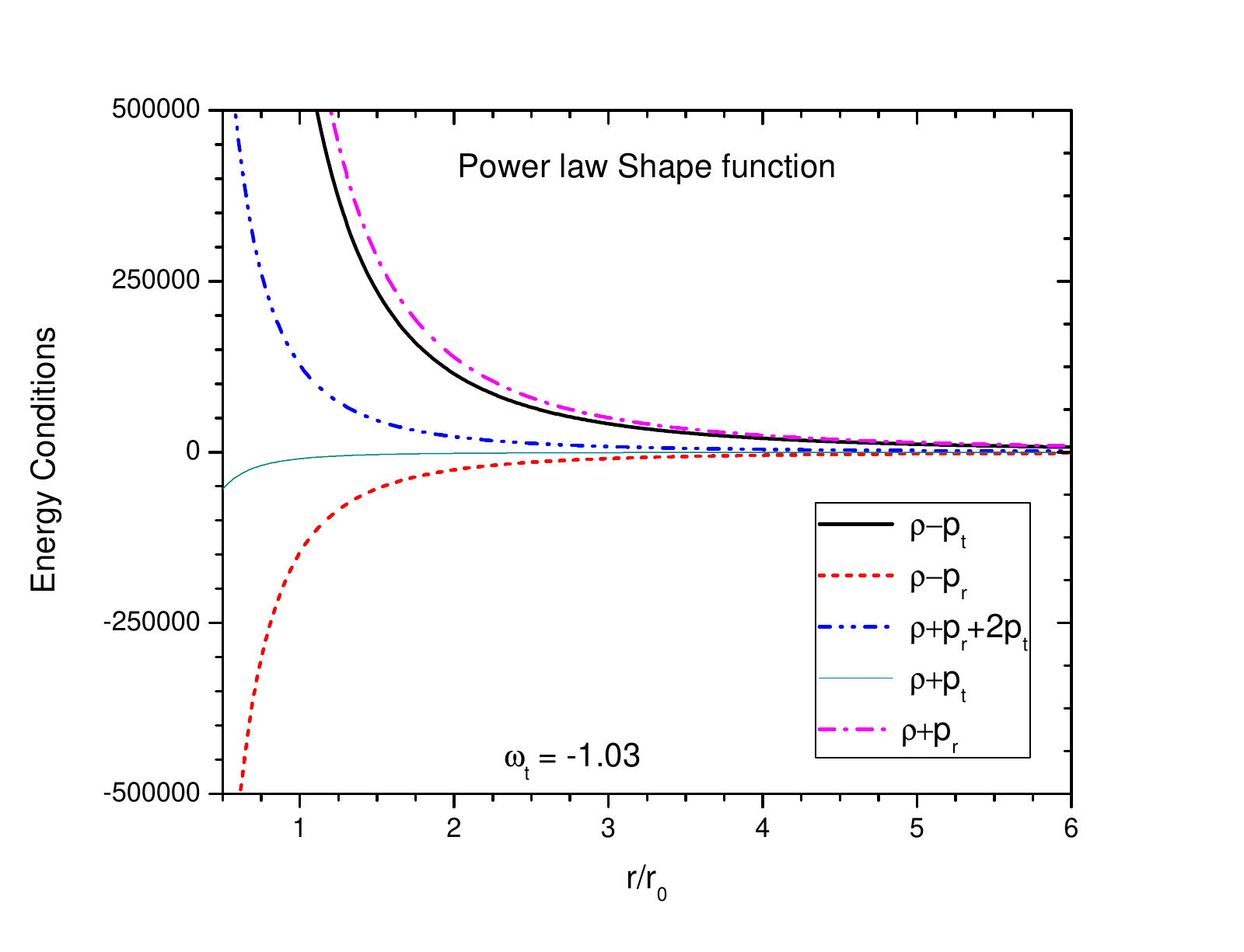}
\endminipage\hfill
\minipage{0.50\textwidth}
\includegraphics[width=\textwidth]{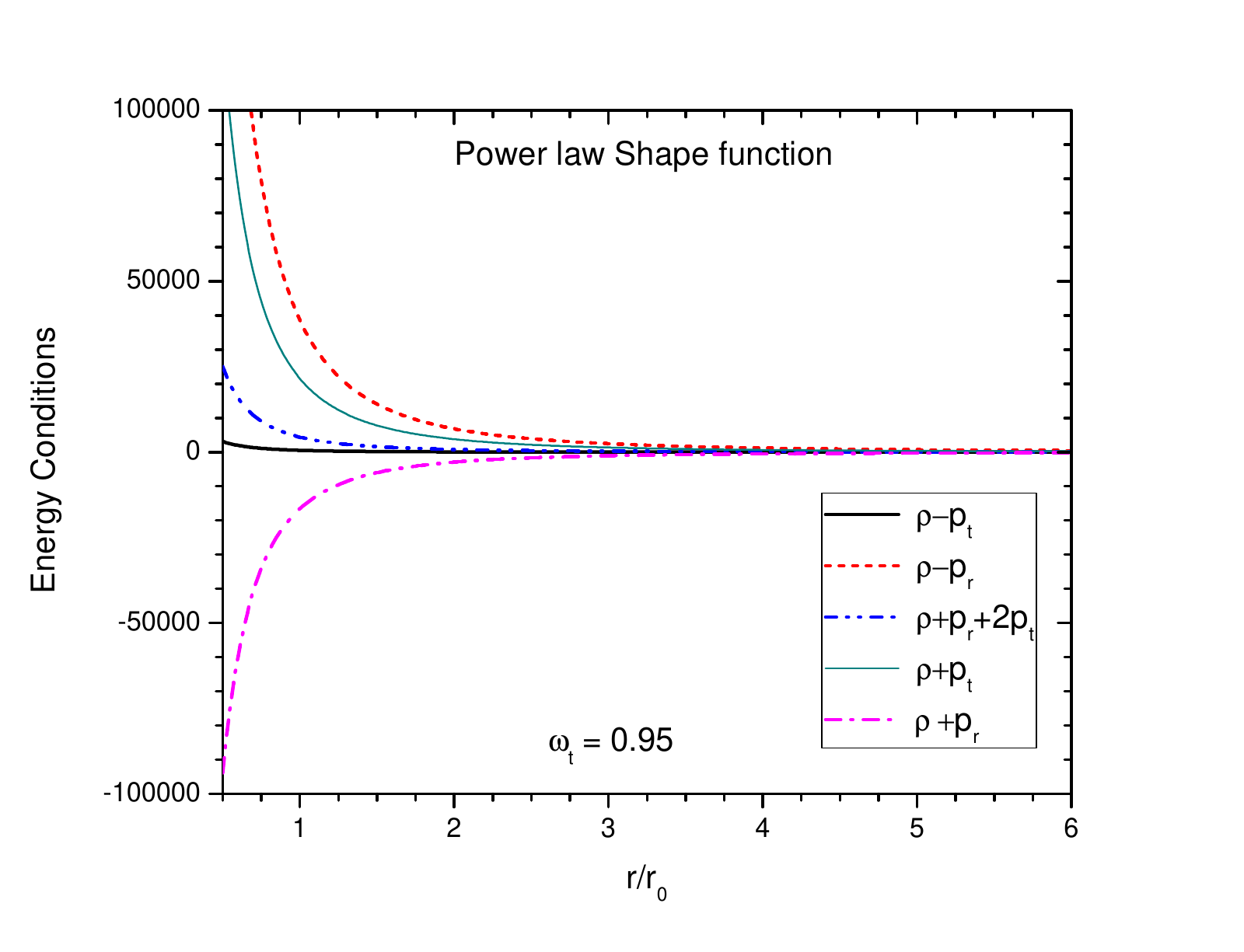}
\endminipage\hfill
\caption{Energy Conditions for the traversable wormholes with power law shape function for $\omega=-1.03$ (Left Panel) and for $\omega=0.95 $ (Right Panel).}\label{FIG.7}
\end{figure*}

\begin{figure}[H]
\centering
\includegraphics[width=0.5\textwidth]{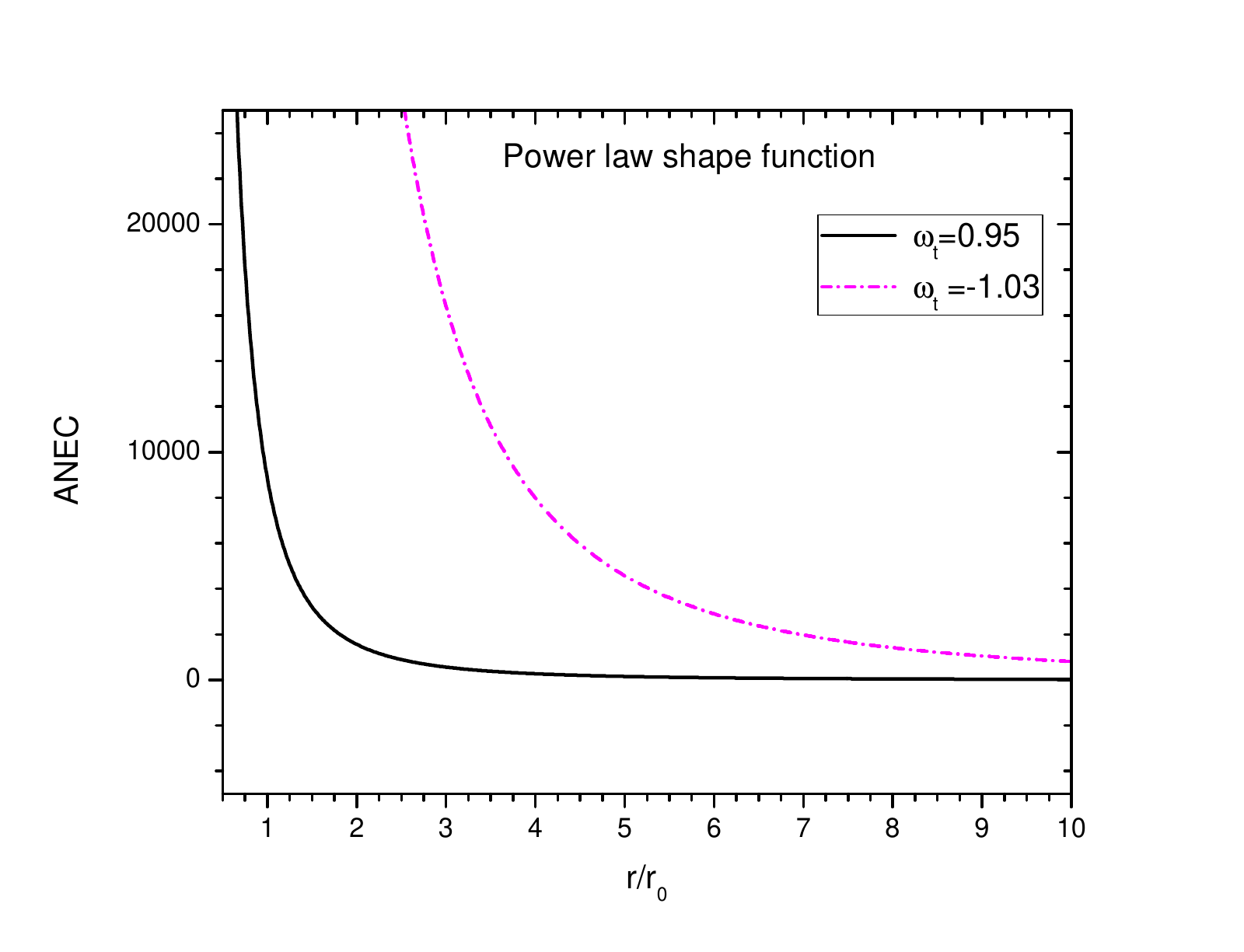}
\caption{ANEC for the Power law shape function.}\label{FIG.8}
\end{figure}

The ANEC is shown in FIG.8 which shows that for both the choices of $\omega_t$, the wormhole matter content satisfies ANEC favouring non-exotic matter wormholes. In this case also, we have investigated with a range of EoS parameter to find the exact range of the EoS parameter for which both the NEC1 and NEC2 are satisfied simultaneously. Our results are shown in TABLE-II. Complete violation of the null energy conditions occur for the range $-0.36 \leq \omega_t \leq 0.19$ and a complete satisfaction of the null energy conditions occur for the range $-1 \leq \omega_t \leq -0.37$. It is not possible to get non-exotic matter wormholes within this geometry and gravity theory for the EoS parameter $\omega_t$ below $-1$.
\begin{table}
\caption{The range of parameter for $\omega_t$ for which NEC1, NEC2 and ANEC are satisfied or violated for the power law shape function. A violation of the energy condition is denoted through ($\times$) and their satisfaction by ($\surd$).}
\begin{tabular}{c|c|c|c}
\hline
range of $\omega_t$ & NEC1      & NEC2      & ANEC       \\
\hline
1					& $\surd$	& $\times$	& $\times$  \\
0.99- 0.72		    & $\times$	& $\surd$	&   $\surd$	\\
0.71 - $0.25$		& $\surd$	& $\times$	&	$\times$	 \\
$0.19$ - $-0.36$	& $\times$	&  $\times$	& $\times$ \\
$-0.37$ - $-1$		& $\surd$ 	& $\surd$	&$\surd$ \\
$-1.01$- $-1.1$	    & $\surd$ 	& $\times$	&$\surd$ \\
\hline
\end{tabular}
\end{table}
\section{Exotic matter content of the traversable wormhole}\label{sec:IV}
Within classical relativity, it is usual that, the matter content of the traversable wormholes violate the null energy condition including the violation of NEC implied from some quantum effects. However, within the purview of geometrically extended gravity theories, it is possible to have smooth wormhole solutions without violating the null energy conditions. In the present study, we have shown that, the possibility of the existence of non-exotic matter wormholes occur for certain range of the EoS parameter chosen. In view of this, it is required to estimate the exotic matter content of the wormhole.

The exotic matter content of the wormhole may be obtained as \cite{Visser2003,Tripathy2021}
\begin{eqnarray}
m  &=& \oint (\rho+p_r)~dV,\\
   &=& 8\pi\int_{0}^{\infty}~(\rho+p_r)~r^2~dr.\label{eq:39}
\end{eqnarray}
\textit{Case-I:}\\
For the exponential shape function, we may write the NEC1 as 
\begin{equation}
\rho+p_r= \left(Ar+B\right)(2\omega_t+\chi-1)e^{2r_0}r^{-2}e^{-2r},
\end{equation}
where 
\begin{eqnarray}
A &=& \frac{2(3-11\omega_r+2\omega_t)}{(2\omega_r\omega_t+2\omega_t+\omega_r^2+3)(8\pi+3\lambda_1)-12\lambda_1+32\pi\omega_r} \nonumber \\
\end{eqnarray}
\begin{eqnarray}
B &=& \frac{12(\omega_r+1)}{(2\omega_r\omega_t+2\omega_t+\omega_r^2+3)(8\pi+3\lambda_1)-12\lambda_1+32\pi\omega_r}.\nonumber \\
\end{eqnarray}
Evaluation of the integral \eqref{eq:39} for the exponential shape function yields
\begin{equation}
m= 2\pi e^{2r_0}(A+2B)(2\omega_t+\chi-1).
\end{equation}
It is straightforward to calculate the exotic matter content for the present model with a given value of the EoS parameter $\omega_t$.\\

\textit{Case-II:}\\
For the power law shape function, we may write the NEC1 as 
\begin{equation}
\rho+p_r= Cr^{-5/2},
\end{equation}
where 
\begin{widetext}
\begin{eqnarray}
C = \frac{2\omega_t+\chi-1}{2}\left[\frac{27+13\omega_r+2\omega_t}{(2\omega_r\omega_t+2\omega_t+\omega_r^2+3)(8\pi+3\lambda_1)-12\lambda_1+32\pi\omega_r}\right]\sqrt{r_0}.
\end{eqnarray}    
\end{widetext}

The exotic matter content of the wormhole with a power law geometry may be obtained through the evaluation of the integral

\begin{equation}
m= 8\pi C\int_{0}^{\infty}~r^{-1/2}~dr, 
\end{equation}
which may be useful for a wormhole whose field only deviates little from the equivalent Schwartzchild one in a region from the throat to certain radial distance $\mathcal {R} \approx r_0+\epsilon$ for $\epsilon <1$.

The exotic matter content for the present case becomes
\begin{equation}
m \approx 16\pi C\left[r_0^{1/2}+\frac{\epsilon}{2r_0^{1/2}}-\frac{\epsilon^2}{8r_0^{3/2}}\right].
\end{equation}

\section{Summary and Conclusion}\label{sec:V}
 In the present work, we have presented traversable wormholes solutions for two different geometries within the set up of an extended gravity theory that involves a term proportional to the square of the energy-momentum tensor. Geometrically modified gravity theories lead to the violation of the energy-momentum conservation which alters the null energy conditions for the matter content of the wormholes. In otherwords, the extended gravity theories may provide ways for the existence of non-exotic matter traversable wormholes. Within the considered squared trace extended gravity theory, we have explored the possibility of the existence of non-exotic matter wormholes.

The structure of the field equations derived for the traversable wormholes put certain constraints on the model parameters also on the EoS parameters considered. In the extended gravity theory, we have considered two model parameters one ($\lambda_1$) signifies the coefficient of the term proportional to the trace of the energy-momentum tensor and the other ($\lambda_2$) to the coefficient of the term proportional to the square of the trace of $T$. It is interesting to note that, the energy density derived for the traversable wormholes does not depend on $\lambda_2$. This does not necessarily imply the ignorance of the squared trace term or of consideration of $\lambda_2=0$.

We have considered two shape functions namely an exponential and a power law ones that satisfy all the conditions required for a viable traversable wormhole geometry. The energy conditions such as the NEC1, NEC2 and SEC are obtained for these shape functions considering some specific EoS parameter within the range $-1.1\leq \omega_t\leq 1$. As has been constrained from the structure of the field equations, the other EoS parameter corresponding to the radial pressure equation is obtained from $\omega_t$. However, the behaviour of the energy conditions for the two geometry considered are shown for $\omega_t=-1.03$ and $\omega_t=0.95$ which provide $\omega_r=-2.66$ and $-2.3$ respectively. For the exponential shape function, the energy density and pressure are obtained to be positive upto $r=0.97r_0$ whereas the tangential pressure is negative for $r=0.97r_0$. On the other hand for $\omega_t=0.95$, the tangential pressure and energy density are obtained to have negative values and the radial pressure is positive throughout the radial distance considered in the work. For the power law shape function, the energy density remains positive for both the choices of the EoS parameter $\omega_t$. The tangential pressure is positive and the radial pressure is negative for $\omega_t=0.95$ and reverse for $\omega_t=-1.03$. However, all of these parameter evolve to assume null values for large radial coordinates.

The prime objective of the present work is to investigate the possibility of the existence of non-exotic matter traversable wormholes. Such wormholes should not violate the null energy conditions NEC1 and NEC2. For the choices of the EoS parameter $\omega_t$ namely $-1.03$ and $0.95$, we could not obtain a situation for both the shape functions where both the null energy conditions are satisfied simultaneously. Also, we calculated the ANEC defined through $\rho+p$ and could not confirm the existence of non-exotic matter wormholes. In view of this, we have varied the choice of the EoS parameter within the range $-1.1\leq \omega_t\leq 1$ and searched for the range of $\omega_t$ for which all the null energy conditions: NEC1, NEC2 and ANEC are satisfied simultaneously and reported the results in TABLE-I and TABLE-II. For the exponential shape function, the NECs are satisfied for the range $-0.37\leq \omega_t\leq -0.44$ for  a wider range of the radial distance and for the power law shape function, non-exotic wormholes are possible for a wider range of the EoS parameter $-0.37\leq \omega_t\leq -1$. One may have non-exotic wormholes for the exponential shape function within the given squared trace gravity model below the specified range but, they may not be viable beyond certain radial distance. If we consider an overlapping range of the EoS parameter for which non-exotic wormholes may exist for both the choices of the shape functions, then we have a specific narrow range $-0.37\leq \omega_t\leq -0.44$. 

We may remark here that, the existence of non-exotic wormholes within the squared trace extended gravity theory is not obvious. It may depend on the choice of the geometry and the shape function. Also, we have a narrow range of allowed EoS parameters for which the existence of non-exotic wormholes can be established. 
\section*{Acknowledgement}
BM and SKT acknowledge the support of IUCAA, Pune (India) through the visiting associateship program.

\end{document}